\documentclass[RNAAS]{aastex62}
\usepackage{url,amssymb,amsmath}

\graphicspath{{./}{figures/}}

\begin{document}

\title{On the progenitor system of V392 Persei}

\email{M.J.Darnley@ljmu.ac.uk}

\author[0000-0003-0156-3377]{M. J. Darnley}
\affil{Astrophysics Research Institute, Liverpool John Moores University, IC2 Liverpool Science Park, Liverpool, L3 5RF, UK}

\author[0000-0002-1359-6312]{S. Starrfield}
\affil{School of Earth and Space Exploration, Arizona State University, Tempe, AZ 85287-1504, USA}

\received{2018 May 2}
\accepted{2018 May 2}
\revised{2018 May 6}
\published{TBC}


\section{}
On 2018 April 29.474\,UT \citet{Discovery} reported the discovery of a new transient, TCP~J04432130+4721280, at $m=6.2$ within the constellation of Perseus.  Follow-up spectroscopy by \citet{LeadbeaterSpec} and \citet{WagnerATel} independently verified the transient as a nova eruption in the `Fe\,{\sc ii} curtain' phase; suggesting that the eruption was discovered before peak. \citet{Peak} reported that the nova peaked on April 29.904 with $m=5.6$. \citet{Discovery} noted that TCP~J04432130+4721280 is spatially coincident with the proposed Z~Camelopardalis type dwarf nova (DN) V392~Persei \citep[see][]{1993PASP..105..127D}. V392~Per is therefore among just a handful of DNe to subsequently undergo a nova eruption \citep[see, e.g.,][]{2016Natur.537..649M}.

The AAVSO\footnote{\url{http://aavso.org}} 2004--2018 light curve for V392~Per indicates a quiescent system with $V\sim16-17$\,mag, punctuated with three of four DN outbursts, the last in 2016. \citet{1993PASP..105..127D} recorded a quiescent range of $15.0\leq m_\mathrm{pg}\leq17.5$, \citet{1994A&AS..107..503Z} reported a magnitude limit of $V>17$. These observations suggest an eruption amplitude of $\lesssim12$ magnitudes, which could indicate the presence of an evolved donor in the system.

The eruption spectroscopy indicated relatively high ejecta velocities ($\sim5000$\,km\,s$^{-1}$) for a classical nova (CN), with the H$\alpha$ profile possibly containing extended -- even higher velocity -- emission around the central peak \citep[see the spectrum contained within][]{WagnerATel}. Such high velocities, coupled with a low eruption amplitude, and also the prompt post-eruption detection of $\gamma$-ray emission \citep{GammaRay}, are features one might expect to see from a recurrent nova (RN), particularly one within a symbiotic binary.

The following photometry of the quiescent V392~Per is contained within the 2MASS All-Sky Catalog of Point Sources \citep{2003yCat.2246....0C} and {\it WISE} All-Sky Source Catalog \citep{2012yCat.2311....0C}:\ $J=13.766\pm0.031$, $H=13.290\pm0.038$, $K_\mathrm{S}=13.062\pm0.037$, $w1=12.878\pm0.030$ (3.3\,$\mu$m), $w2=12.761\pm0.032$ (4.6\,$\mu$m), the system was not detected in {\it WISE} bands 3 and 4 (12 and 22\,$\mu$m).

{\it Gaia} Data Release 2 \citep[DR2;][]{2016A&A...595A...1G,2018arXiv180409365G} contains a parallax measurement for V392~Per of $0.257\pm0.052$\,mas, which {\it could} indicate a distance of $3.9^{+1.0}_{-0.6}$\,kpc. We note the caveats regarding DR2 distance determinations \citep[see][]{2018arXiv180409366L}, particularly those regarding unresolved binary systems.  The 3D dust maps of \citet{2015ApJ...810...25G,2018arXiv180103555G} yield a reddening of $E_{\mathrm{B-V}}=0.9\pm0.1$ over the {\it Gaia} distance range. Taking both this distance and reddening at face value, the absolute magnitude of the eruption could have reached $M_V=-9.5^{-0.8}_{+0.7}$ (or $-10.1^{-0.8}_{+0.7}$ assuming a peak of $m_V=5.6$) -- in either case, this could be inherently a very luminous eruption.

\begin{figure}
\centering\includegraphics[width=0.696\textwidth]{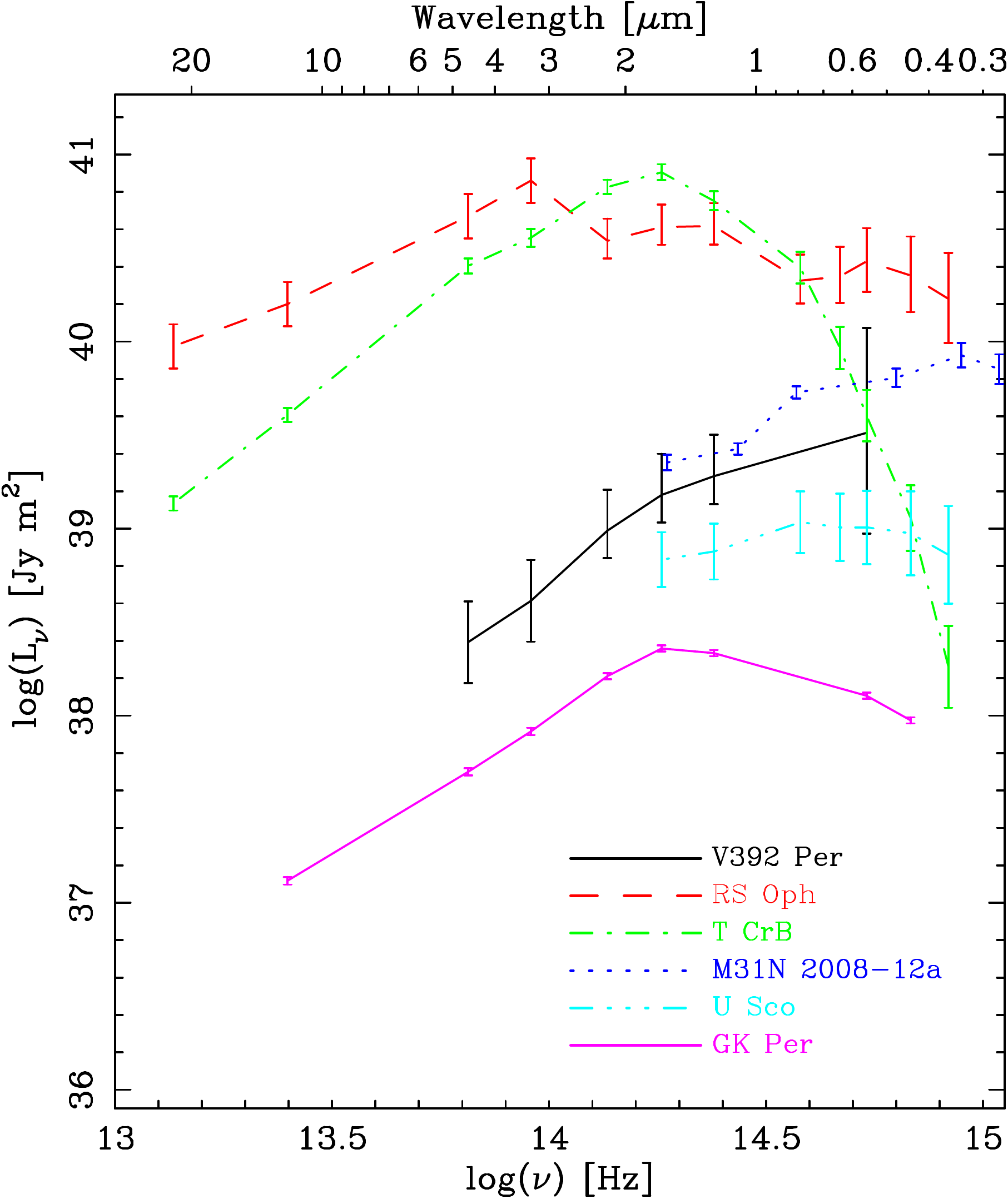}
\caption{Distance and extinction corrected quiescent SEDs of V392~Per, RS~Oph, T~CrB, M31N~2008-12a, U~Sco, and GK~Per. The error bars include photometric, distance, and extinction uncertainties. The lines are to aid the reader.\label{sed.fig}}
\end{figure}

Figure~\ref{sed.fig} shows the quiescent spectral energy distribution (SED) of V392~Per using {\it WISE} and 2MASS data, the {\it Gaia} distance, and extinction as above. The quiescent SED is compared to the RNe RS~Ophiuchi, T~Coronae Borealis, M31N~2008-12a, and U~Scorpii, and that of the CN, DN, and intermediate polar, GK~Persei, using data within \citet{2012ApJ...746...61D,2017ApJ...849...96D} and \citet{2014MNRAS.444.1683E}. We utilise the {\it Gaia} distances for all objects (except U~Sco and M31N~2008-12a), these are consistent with those recorded in \citet{2012ApJ...746...61D}. As noted by \citet{2014MNRAS.444.1683E}, the {\it WISE} photometry of U~Sco may be affected by strong emission lines and therefore is not included.

In conclusion, even after correcting for the large {\it Gaia} distance and the large extinction, the SED of the V392~Per progenitor is not consistent with the system containing a red giant/symbiotic donor.  However, the SED appears similar to those of U~Sco and GK~Per, a RN and CN, respectively, which contain sub-giant donors, or even that of M31N~2008-12a with it's proposed low luminosity giant or `red clump' donor. V392~Per is unlikely to be more distant than implied by {\it Gaia} -- but if significantly closer, the photometry could be consistent with a main sequence donor.

\acknowledgments

With thanks to Brad Schaefer and Patrick Schmeer. This publication makes use of data products from the Wide-field Infrared Survey Explorer, which is a joint project of the University of California, Los Angeles, and the Jet Propulsion Laboratory/California Institute of Technology, funded by the National Aeronautics and Space Administration (NASA). This publication makes use of data products from the Two Micron All Sky Survey, which is a joint project of the University of Massachusetts and the Infrared Processing and Analysis Center/California Institute of Technology, funded by NASA and the National Science Foundation. This work has made use of data from the European Space Agency (ESA) mission {\it Gaia} (\url{https://www.cosmos.esa.int/gaia}), processed by the {\it Gaia} Data Processing and Analysis Consortium (DPAC, \url{https://www.cosmos.esa.int/web/gaia/dpac/consortium}). Funding for the DPAC has been provided by national institutions, in particular the institutions participating in the {\it Gaia} Multilateral Agreement.


\begin{thebibliography}{}
\bibitem[Buczynski(2018)]{Peak} Buczynski, D. 2018, CBAT, IAU, \url{http://www.cbat.eps.harvard.edu/unconf/followups/J04432130+4721280.html}
\bibitem[Cutri et al.(2003)]{2003yCat.2246....0C} Cutri, R.~M., Skrutskie, M.~F., van Dyk, S., et al.\ 2003, VizieR Online Data Catalog , II/246.
\bibitem[Cutri et al.(2012)]{2012yCat.2311....0C} Cutri, R.~M., et al.\ 2012, VizieR Online Data Catalog , II/311.
\bibitem[Darnley et al.(2012)]{2012ApJ...746...61D} Darnley, M.~J., Ribeiro, V.~A.~R.~M., Bode, M.~F., Hounsell, R.~A., \& Williams, R.~P.\ 2012, \apj, 746, 61 
\bibitem[Darnley et al.(2017)]{2017ApJ...849...96D} Darnley, M.~J., Hounsell, R., Godon, P., et al.\ 2017, \apj, 849, 96 
\bibitem[Downes \& Shara(1993)]{1993PASP..105..127D} Downes, R.~A., \& Shara, M.~M.\ 1993, \pasp, 105, 127 
\bibitem[Evans et al.(2014)]{2014MNRAS.444.1683E} Evans, A., Gehrz, R.~D., Woodward, C.~E., \& Helton, L.~A.\ 2014, \mnras, 444, 1683 
\bibitem[Gaia Collaboration et al.(2016)]{2016A&A...595A...1G} Gaia Collaboration, Prusti, T., de Bruijne, J.~H.~J., et al.\ 2016, \aap, 595, A1 
\bibitem[Gaia Collaboration et al.(2018)]{2018arXiv180409365G} Gaia Collaboration, Brown, A.~G.~A., Vallenari, A., et al.\ 2018, arXiv:1804.09365 
\bibitem[Green et al.(2015)]{2015ApJ...810...25G} Green, G.~M., Schlafly, E.~F., Finkbeiner, D.~P., et al.\ 2015, \apj, 810, 25 
\bibitem[Green et al.(2018)]{2018arXiv180103555G} Green, G.~M., Schlafly, E.~F., Finkbeiner, D., et al.\ 2018, arXiv:1801.03555 
\bibitem[Leadbeater(2018)]{LeadbeaterSpec} Leadbeater, R. 2018, ARAS Spectroscopy Forum, \url{http://www.spectro-aras.com/forum/viewtopic.php?f=5&t=2015}
\bibitem[Li et al.(2018)]{GammaRay} Li, K.-L., Chomiuk, L., Strader, J. 2018, ATel, 11590
\bibitem[Lindegren et al.(2018)]{2018arXiv180409366L} Lindegren, L., Hernandez, J., Bombrun, A., et al.\ 2018, arXiv:1804.09366 
\bibitem[Mr{\'o}z et al.(2016)]{2016Natur.537..649M} Mr{\'o}z, P., Udalski, A., Pietrukowicz, P., et al.\ 2016, \nat, 537, 649 
\bibitem[Nakamura(2018)]{Discovery} Nakamura, Y. 2018, CBAT, IAU, \url{http://www.cbat.eps.harvard.edu/unconf/followups/J04432130+4721280.html}
%\bibitem[Schlegel et al.(1998)]{1998ApJ...500..525S} Schlegel, D.~J., Finkbeiner, D.~P., \& Davis, M.\ 1998, \apj, 500, 525 
\bibitem[Wagner et al.(2018)]{WagnerATel} Wagner, R.~M., Terndrup, D., Darnley, M.~J., et al.\ 2018, ATel, 11588 
\bibitem[Zwitter \& Munari(1994)]{1994A&AS..107..503Z} Zwitter, T., \& Munari, U.\ 1994, \aaps, 107, 503 



\end{thebibliography}
\end{document}